\documentclass[onecolumn,amsmath,amssymb,floatfix,pra,shownopacs,footinbib,superscriptaddress]{revtex4-1}

\ifx
\pdfoutput\undefined
 \usepackage[dvips]{graphicx}
 \else
 \usepackage[pdftex]{graphicx}
 \fi

\usepackage{epstopdf}
\usepackage{amsmath,amssymb,natbib,bm,graphicx,url}
\usepackage{graphicx}
\usepackage{psfrag}
\usepackage{amssymb}
\usepackage{subfigure}
\usepackage{amsthm}
\usepackage{color}
\usepackage{wasysym}

\newcommand{\Eq}[1]{Eq.~(\ref{#1})}

\newcommand{\Sec}[1]{Sec.~\ref{#1}}
\newcommand{\Fig}[1]{Fig.~\ref{#1}}

\newcommand{\be}{\begin{equation}}
\newcommand{\ee}{\end{equation}}
\newcommand{\bea}{\begin{eqnarray}}
\newcommand{\eea}{\end{eqnarray}}

\newcommand{\bra}[1]{\left\langle \, #1 \right|}
\newcommand{\ket}[1]{\left| #1  \right\rangle}

\newcommand{\dagg}[1]{#1^\dagger}

\newcommand{\im}{\text{Im}}
\newcommand{\re}{\text{Re}}

\begin{document}

\title{Stochastic Master Equation Analysis of Optimized Three-Qubit Nondemolition Parity Measurement}

\author{L. Tornberg}
\affiliation{Chalmers University of Technology, SE-41296 Gothenburg, Sweden}
\author{Sh. Barzanjeh}
\affiliation{Institute for Quantum Information, RWTH Aachen University, D-52056 Aachen, Germany}
\author{David P. DiVincenzo}
\affiliation{Institute for Quantum Information, RWTH Aachen University, D-52056 Aachen, Germany}
\affiliation{Peter Gr{\"u}nberg Institute: Theoretical Nanoelectronics, Research Center J{\"u}lich}
\affiliation{J{\"u}lich-Aachen Research Alliance (JARA), Fundamentals of Future Information Technologies}
\begin{abstract}
We analyze a direct parity measurement of the state of three superconducting qubits in circuit quantum electrodynamics. The parity is inferred from a homodyne measurement of the reflected/transmitted microwave radiation and the measurement is direct in the sense that the parity is measured without the need for any quantum circuit operations or for ancilla qubits.  Qubits are coupled to two resonant cavity modes, allowing the steady state of the emitted radiation to satisfy the necessary conditions to act as a pointer state for the parity. However, the transient dynamics violates these conditions and we analyze this detrimental effect and show that it can be overcome in the limit of weak measurement signal. Our analysis shows that, with a moderate degree of post-selection, it is possible to achieve post-measurement states with fidelity of order 95\%. We believe that this type of measurement could serve as a benchmark for future error-correction protocols in a scalable architecture. 
\end{abstract}
\date{\today}
\maketitle
%
\section{Introduction}
It is now well established that there are many different ways of achieving, within circuit quantum electrodynamics (cQED), the essential primitive operations for quantum information processing.  Beyond protocols for achieving highly accurate single-\cite{bench1} and two-qubit\cite{bench2} gate operations, the achievement of fast, flexible, accurate quantum measurements\cite{ssr} is also essential.  The current experimental emphasis on reliable feedback of measurement data to control subsequent quantum operations\cite{goodM1, goodM2, goodwithF} fulfils the theoretical hope that such capabilities will find important uses in reliable quantum information processing\cite{LT12}.  In particular, the achievement of successful fault tolerant quantum computation relies on the implementation of adaptive gate sequences conditioned on the results of specific kinds of quantum measurements\cite{BT}, namely multi-qubit parity measurements\cite{LT10}.  In a parity measurement,  the measurement outcome is to be one single bit, regardless of the number of qubits involved.  The bit should simply indicate whether the number of ones in the set of measured qubits is even or odd.  It is essential for the proper functioning of this measurement that no other information about the qubits be uncovered by the measurement.  Furthermore, it is necessary, for applications in quantum error correction, that the measurement be ``quantum non-demolition'' (QND); if the state of the qubits before measurement is an eigenstate of the measurement (i.e., is purely even or purely odd), then the final quantum state of the qubits should be unchanged.\\

In the preferred (topological) error correction code schemes, the parity of four\cite{BT} or three\cite{3par} qubits is needed.  It has generally been assumed that this parity measurement would be accomplished with a quantum circuit involving one- and two-qubit gates; the parity is thus computed by elementary logic operations, with the result stored in another ancillary qubit.  The parity measurement is then completed by a conventional single-qubit measurement on the ancilla.  This measurement does not even have to have QND character on the ancilla -- the net result is a QND measurement on the data qubits.  Nevertheless, there are reasons for wishing to replace this quantum circuit with a more direct procedure.  First, a direct parity measurement dispenses with the need for extra ancilla qubits.  Furthermore, the problem of the accumulation of error is ameliorated.  That is, in the circuit approach, the net error will be the sum of the errors of each of the quantum gates and of the single-qubit measurement.  This is no fewer than four individual operations; it is known that for achievement of fault tolerance, each of these individual operations needs to have an error rate no larger than about 1\%\cite{goodest}.   This means that the ``all in one'' operation studied here is permitted to have a larger error rate, around 4\%, say.\\

Several detailed concepts for direct two-qubit parity measurements have been  analyzed recently\cite{mabuchiway,coherent2,LT10,LT12,NG}, with a number of them being promising for applications.  The central idea of these approaches is that qubits are off-resonantly (i.e. dispersively) coupled to a cavity mode; the frequency of the mode is shifted by an amount dependent on the state of the qubits, and this shift is then read out by measuring the phase of a microwave tone either transmitted through or reflected from the cavity.   \\

For the most part, these earlier proposals have no straightforward generalisation to measurement of the parity of more than two qubits.  Refs. \cite{mabuchiway,firat} indicated that a generalisation of previous schemes that would make multiqubit parity measurement possible involves {\em multiple resonant modes}.  In particular \cite{firat} showed that, by having just two cavity modes, each of which is subject to qubit-state-dependent dispersive shifts, three qubit parity measurements become possible.  The recent proposal of Nigg and Girvin \cite{NG} is clearly extendible to multi-qubit parity measurement; by loading a cavity with a coherent state in a precisely timed way, the state's phase can accumulate information about a particular subset of qubits (with others removed by refocussing), such that the subsequent dispersive measurement of another ancilla qubit can give a readout of any subset parity.\\

In this paper, we take up a detailed analysis of the multi-qubit parity measurement proposal of \cite{firat} using the stochastic master equations used to represent realistic homodyne measurements, as in Refs \cite{LT10,LT12,coherent2}.  \cite{firat} developed the two-mode concept using a completely different approach, which involved obtaining scattering parameters from a classical linear electrical circuit analysis, combined with an unrealistic model of measurement in which a hypothetical von-Neumann measurement is performed instantaneously after a coherent-state probing pulse has completed its unitary interaction with the system.  As a part of the present work, we provide a new derivation based on input-output theory\cite{MW} of the quantum optics of a cavity, coupled to a continuum, with two closely spaced resonant modes. \\

Many of the broad features of \cite{firat} are confirmed in the present realistic study: there exists a choice of parameters, in which all the relevant parameters of the problem (the dispersive coupling of qubit to cavity, the detuning of the two cavity modes from each other, and the coupling strength of the two modes to the continuum) are of comparable strength, for which a successful three-qubit parity measurement is obtained. In fact, our present analysis provides new, simple formulas for the ideal setting of all these parameters.  For these settings, the steady-state statistics of the homodyne measurement are identical for any state in one of the parity subspaces (even or odd).  The transient response, however, does distinguish individual states, and thus degrades the fidelity of the parity measurement.  While \cite{firat} indicated that a good strategy for dealing with these transient effects is to use a low-intensity, weak measurement of long duration, the details of the present optimisation of the measurement in light of the transient effects were not anticipated by the earlier analysis.  Furthermore, in the present study we consider a realistic measurement setting in which qubit decay, determined fundamentally in the cavity setting by the Purcell effect\cite{Pur}, constrains the improvement that can be obtained by prolonging the measurement.  Our optimisations indicate that despite the current limitations of superconducting qubit-cavity systems, parity measurements of impressive fidelity (c. 95\%) will be possible, but only if we permit a moderate degree (c. 50\%) of post-selection to retain only those cases where the homodyne measurement is most conclusive.  Better performance with the presently-analyzed scheme is not precluded, but would appear to require qubits with longer $T_1$ and $T_2^*$ times.\\

The paper is organised as follows. In \Sec{sec:system}, we present the model for the circuit QED system, containing two resonant modes and three qubits. By tracing out the mode degrees of freedom we derive an effective stochastic master equation for the qubit dynamics. In \Sec{sec:paritymeas}, we discuss the desired properties of a parity measurement and derive the optimal values of circuit parameters to obtain these. We define the measures of measurement fidelity and study the effect of measurement inefficiency in \Sec{sec:measfidelity} and \Sec{sec:eta} respectively. \Sec{sec:transients} is devoted to the study of transient effects and a strategy for mitigating the undesired measurement back-action is presented. In \Sec{sec:results} we give the main results and conclude in \Sec{sec:conclusions}.

\section{The system}\label{sec:system}
\begin{figure}[htp] 
\centering
\includegraphics[scale=0.87]{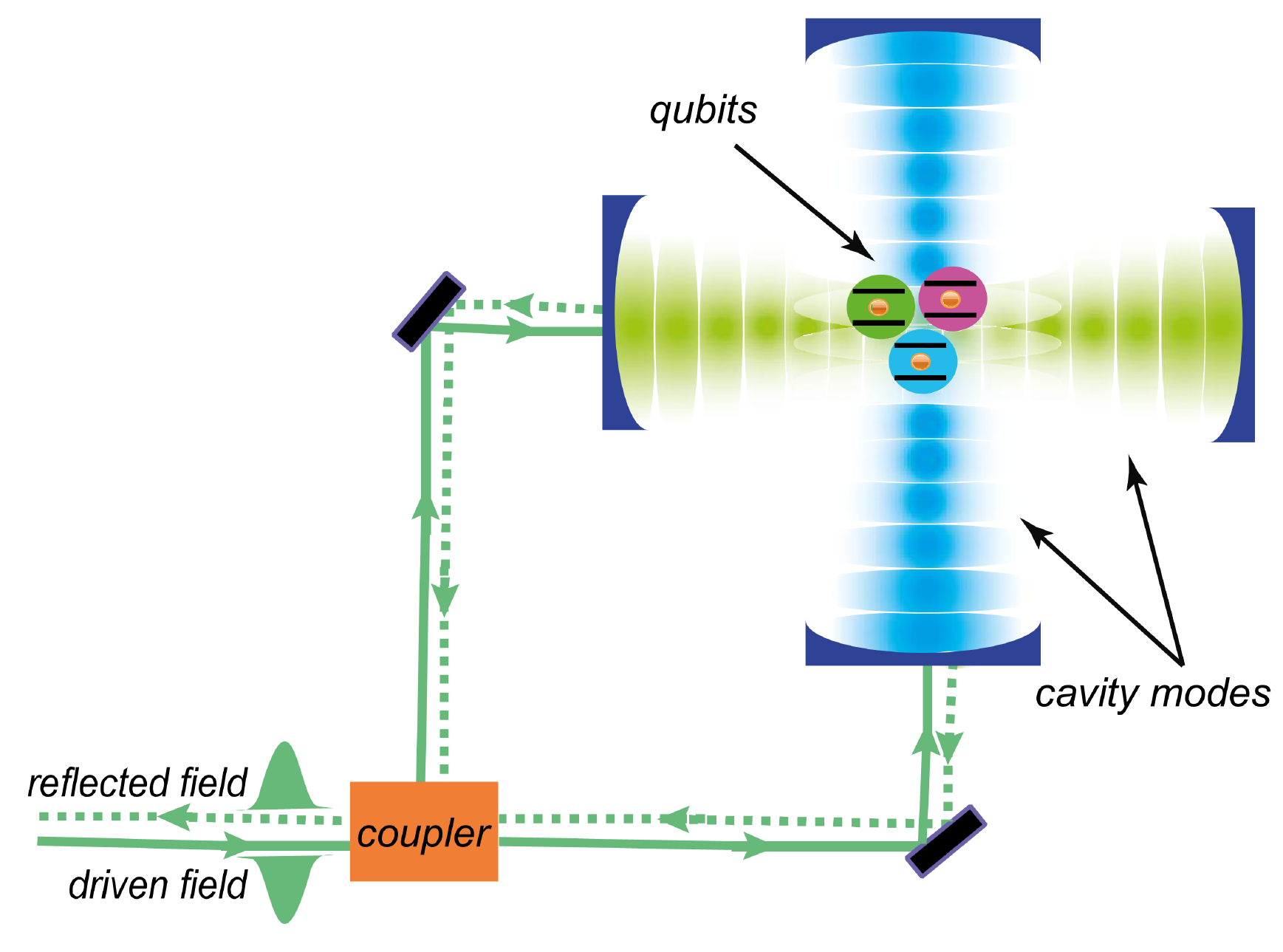}
\caption{A possible physical realization of the three-qubit parity measurement analyzed in this paper.  This concept uses elements from traditional optics and cavity QED; Ref. \cite{firat} illustrates several possible realizations of the measurement using microwave techniques, i.e., using circuit QED.  Three atoms or artificial atoms are held in space, either by trapping techniques or by being embedded in a crystal, in the middle of a crossed-mode double cavity.  The two relevant horizontal and vertical modes are to be slightly detuned from one another, and are far detuned from the atomic transitions, so that the cavity-atom interaction is dispersive.  The two modes are driven simultaneously with pulsed coherent radiation whose frequency is in between that of the two cavity modes.  Parity information is extracted by a homodyne measurement of the reflected field.  The ``fiber coupler'' accomplishing the splitting and combining can be a standard three-port component such as a symmetric Y-branch coupler.   A modification of this setup is straightforwardly possible in which the output field emerges in transmission rather than reflection.}
\label{fig:system}
\end{figure}

The system whose parity we want to measure consists of three (artificial) atoms coupled to two fundamental modes of a cavity (or two different cavities), which couple to a common input and output continuum, as depicted in \Fig{fig:system}. 
For simplicity, we neglect the possible influence of higher qubit levels and approximate each atom as a two-level system. The system is operated in the dispersive regime, where the transition frequency $\Omega_i$ of qubit $i$ is far detuned from the resonance frequency of either resonator mode $\omega_j$ such that $g_i^j/|\Omega_i - \omega_j| \ll 1$ where $g_i^j$ is the coupling strength between qubit $i$ and mode $j$. In this regime the Hamiltonian, in the rotating frame defined by the measurement-tone frequency $\omega_m$, is given by \cite{blaisPRA2004}
\begin{equation}\label{eq:H}
H = \Big(\Delta^b+\sum_{j=1}^{3}\chi^b_j \sigma_z^{(j)}\Big)b^{\dagger}b+\Big(\Delta^a+\sum_{j=1}^{3}\chi^a_j \sigma_z^{(j)}\Big)a^{\dagger}a+\sum_{j=1}^{3}\frac{\Omega_j+\chi^a_j+\chi^b_j}{2}\sigma_z^{(j)} + \Big[\epsilon(t)(\sqrt{\kappa^a}a^{\dagger}+\sqrt{\kappa^b}b^\dagger)+ \mathrm{h.c.}\Big].
\end{equation}
where  $ \Delta^k=\omega^k-\omega_m $ (with $ k=a,b $) are the cavity detunings,  and $ \chi_j^k=(g_j^k)^2/\Delta^k_{j} $ ($j=1,2,3$ and $k=a,b$) are the dispersive coupling strengths with $\Delta_j^{k}=\omega^k-\Omega_j  $. The amplitude of the measurement signal is given by $\epsilon(t)$. The resonator modes are described by the annihilation (creation) operators $ a(a^{\dagger}) $ and  $ b(b^{\dagger}) $. The coupling between resonator mode $i$ and the environment is given by $\kappa_i$. In the absence of measurement, the master equation describing the system evolution is given by
\be\label{eq:master}
\dot{\rho} = -i[H,\rho]+\sum_{j=1}^{3}\Big(\gamma_{1j}\textit{D}[\sigma_{-}^{(j)}]\rho+ \frac{\gamma_{\varphi j}}{2}\textit{D}[\sigma_{z}^{(j)}]\rho\Big)+\textit{D}\Big[\sqrt{\kappa^a}a+\sqrt{\kappa^b}b\Big]\rho+
\kappa^a\sum_{j=1}^{3}\textit{D}\Big[\lambda^a_j\sigma_{-}^{(j)}\Big]\rho+
\kappa^b\sum_{j=1}^{3} \textit{D}\Big[\lambda^b_j\sigma_{-}^{(j)}\Big]\rho,
\ee
where $\textit{D}\left[c\right]\rho = c\rho c^\dagger + 1/2(c^\dagger c \rho + \rho c^\dagger c )$ is a dissipation superoperator of Lindblad form \cite{lindbladCMP1976} and $\gamma_{1j}$ and $\gamma_{\varphi j}$ are the relaxation and dephasing rates of qubit $j$ respectively. The last two terms describe the Purcell relaxation \cite{Pur} where $ \lambda_j^k=g_j^k/\Delta^k_{j} $ ($j=1,2,3$ and $k=a,b$) and we have assumed distinct qubit frequencies such that $|\Omega_i - \Omega_j| \gg \kappa^a \lambda^a_j \lambda^a_i$ which allows us to neglect all cross-terms between operators belonging to different qubits. This assumption is also important if we want to suppress the direct coupling between qubits mediated by virtual photons \cite{blaisPRA2004}. 

In such a two-mode setting, it would be common to also have terms in the Hamiltonian involving mode-mode coupling, i.e., terms containing $a^\dagger b$.  While such terms are indeed typically nonzero, it has been shown that, using the flexibility offered within circuit QED, circuit structures can readily be devised where these interactions are tuned away\cite{tunecQED}. While such terms would not fundamentally alter the parity-measurement scheme that we analyze here, we find that the study of the effects are more transparent with the minimal Hamiltonian Eq. (\ref{eq:H}), which we will henceforth employ throughout this paper.

From the point of view of the cavity degrees of freedom, \Eq{eq:H} and \Eq{eq:master} describe the generation and evolution of coherent states, whose amplitudes are governed by the differential equations \cite{gambettaPRA2006}
\begin{eqnarray}\label{eq:cavity_amplitudes}
\dot{\alpha}_{ijk}&&=-i\sqrt{\kappa^a}\epsilon-i(\Delta^a+\chi^a_{ijk})\alpha_{ijk}-\frac{\kappa^a}{2}\alpha_{ijk}-\frac{\sqrt{\kappa^a\kappa^b}}{2}\beta_{ijk},\nonumber \\
\dot{\beta}_{ijk}&&=-i\sqrt{\kappa^b}\epsilon-i(\Delta^b+\chi^b_{ijk})\beta_{ijk}-\frac{\kappa^b}{2}\beta_{ijk}-\frac{\sqrt{\kappa^a\kappa^b}}{2}\alpha_{ijk}.
\end{eqnarray}
such that the cavity fields are entangled with the qubit states through the coupling $\chi^m_{ijk}=\langle ijk|\sum_{l=1}^3 \chi_l^m \sigma_z^{(l) }|ijk\rangle $, with ($ m=a,b $). In this way, the cavity fields act like pointer states with allows us to indirectly infer the state of the qubit system through a measurement on the field.\\

The unconditional evolution described by \Eq{eq:master} is sufficient if one is interested in calculating average quantities of system operators. When studying the performance of a measurement it is however necessary to calculate properties of the system conditioned on a certain subset of measurement results. For this purpose, we need to describe the system evolution conditioned on a single measurement result. In circuit QED, phase sensitive amplification allows for the equivalent of homodyne detection in optics. The system dynamics including the measurement back action is described by the stochastic master equation \cite{wisemanPRA1993}
\begin{equation}\label{eq:SME}
d\rho= \mathcal{L} \rho dt + \sqrt{\eta}\textit{M}\Big[(\sqrt{\kappa^a}a+\sqrt{\kappa^b}b )e^{-i \phi}\Big]\rho dW(t),
\end{equation}
where $\mathcal{L}$ is the superoperator written in \Eq{eq:master} and $\textit{M}[c]\rho =c\rho+\rho c^{\dagger}-\langle c+c^{\dagger}\rangle \rho$ is the superoperator describing the measurement back-action and $\eta$ is the efficiency of the measurement. The stochastic evolution, fundamentally originating from the collapse of the state, is realized through the Wiener process $dW(t)$ with the defining statistical properties $\mathrm{E}[dW(t)] = 0$ and $\mathrm{E}[dW(t)^2] = dt$ \cite{klebaner}. The measurement signal is given by the homodyne current
\be\label{eq:current}
j(t)dt = \sqrt{\eta}\big\langle  ( \sqrt{\kappa^a}a+\sqrt{\kappa^b}b )e^{-i \phi} + (\sqrt{\kappa^a}\dagg{a}+\sqrt{\kappa^b}\dagg{b} )e^{i \phi}  \big\rangle dt + dW(t).
\ee
\Eq{eq:SME} and \Eq{eq:current} can in principle be used to numerically study the evolution of the system and the performance of the measurement. However, to gain qualitative understanding with the long term goal of achieving a high fidelity measurement, it is necessary to reduce these equations such that they describe the evolution of the qubits' degree of freedom only. 
\subsection{Effective stochastic master equation of three-qubit/two-mode circuit QED}
The elimination of the cavity degrees of freedom to obtain an effective SME for the qubits has been treated in detail in \cite{gambettaPRA2008, coherent2}. There the analysis was done for one and two qubits coupled to a single cavity mode. Here we extend this derivation to the case of more cavity modes and qubits. The elimination of cavity degrees of freedom is carried out by moving to a frame of reference which follows the average cavity field, whose state is conditioned on the state of the qubits:
\begin{equation}\label{eq:polaron}
P=\sum_{i,j,k=0,1}D_a(\alpha_{ijk})D_b(\beta_{ijk})\Pi_{ijk},
\end{equation}
where $\alpha$ and $ \beta$ are the coherent amplitudes of cavity modes $ a $ and $ b $, respectively and $D_c(\gamma) = \exp(\gamma c^\dagger - \gamma^*c)$ is the displacement operator with respect to each cavity field \cite{gerryKnight}. $\Pi_{ijk}=|ijk\rangle\langle ijk| $ are projection operators onto the basis states of the three-qubit Hilbert space. The field dynamics in this frame of reference is given by the vacuum fluctuations only and in the limit $\gamma_{1j}\ll \kappa_i$ we can trace out the photonic states, yielding the effective master equation 
\begin{eqnarray}\label{eq:SMEeff}
d \rho&&=-i\Big[\sum_{j=1}^{3}\frac{\omega^{atom}_j+\chi^a_j+\chi^b_j}{2}\sigma_z^{(j)},\rho \Big]dt+\Big(\sum_{j=1}^{3}\gamma_{1j}\textit{D}[\sigma_{-}^{(j)}]+ \frac{\gamma_{\varphi j}}{2}\textit{D}[\sigma_{z}^{(j)}]+
\kappa^a\textit{D}\Big[\lambda^a_j\sigma_{-}^{(j)}\Big]+\kappa^b \textit{D}\Big[\lambda^b_j\sigma_{-}^{(j)}\Big]\Big)\rho dt\nonumber\\
&&+\sum_{ijk,lmn}\Big(\chi^a_{lmn,ijk}[\mathrm{Im}(\alpha^{*}_{ijk}\alpha_{lmn})+i\mathrm{Re}(\alpha^{*}_{ijk}\alpha_{lmn})]\Big)\Pi_{ijk}\rho \Pi_{lmn} dt \nonumber\\
&&+\sum_{ijk,lmn}\Big(\chi^b_{lmn,ijk}[\mathrm{Im}(\beta^{*}_{ijk}\beta_{lmn})+i\mathrm{Re}(\beta^{*}_{ijk}\beta_{lmn})]\Big)\Pi_{ijk}\rho \Pi_{lmn} dt \\
&&+i\frac{\sqrt{\kappa^a\kappa^b}}{2}\sum_{i\neq l}\sum_{j\neq m}\sum_{k\neq n}\Big[\mathrm{Im}(\alpha_{ijk}\beta^{*}_{lmn})+\mathrm{Im}(\beta_{ijk}\alpha^{*}_{lmn})\Big]\Pi_{ijk}\rho \Pi_{lmn} dt \nonumber\\
&&+\frac{\sqrt{\kappa^a\kappa^b}}{2}\sum_{i\neq l}\sum_{j\neq m}\sum_{k\neq n }\Big[\mathrm{Re}(\beta_{ijk}\alpha^{*}_{lmn})+\mathrm{Re}(\alpha_{ijk}\beta^{*}_{lmn})-\mathrm{Re}(\alpha^{*}_{ijk}\beta_{ijk})-\mathrm{Re}(\alpha^{*}_{lmn}\beta_{lmn})\Big]\Pi_{ijk}\rho \Pi_{lmn} dt \nonumber\\
&&+\sqrt{\eta}\textit{M}\Big[ \Pi_{\Sigma}e^{-i\phi}\Big]\rho dW(t),\nonumber
\end{eqnarray}
where $ \chi^m_{ijk,lmn}=\chi_{ijk}^m - \chi_{lmn}^m $ (with $ m=a,b $). Here, we introduce the short-hand measurement operator $\Pi_{\Sigma}=\sum_{i,j,k}\Sigma_{ijk}  \Pi_{ijk}$, where $\Sigma_{ijk} = \sqrt{\kappa^a}\alpha_{ijk} + \sqrt{\kappa^b}\beta_{ijk}$ is the linear combination of the cavity fields visible through the connection port. 
In addition to the system dynamics, the homodyne current is also be expressed in terms of the qubits' degrees of freedom 
\be\label{eq:currentq}
j(t)dt = \sqrt{\eta}\big\langle \Pi_\Sigma e^{-i \phi} + \dagg{\Pi_\Sigma}e^{i \phi}  \big\rangle dt + dW(t). 
\ee

\section{Parity measurement}\label{sec:paritymeas}
\begin{figure}[ht]
\centering
\subfigure[ The solution to \Eq{eq:sigma_ss} as a function of $\kappa^a = \kappa^b$ and $\Delta^a = -\Delta^b$. The blue (red) surfaces show the solution for the negative (positive) parity subspace. The intersection between the planes is indicated below the solution (see text for details). The (optimal) black point shows $\kappa^a = \kappa^b = 2\chi$ and $\Delta^a = -\Delta^b = \sqrt{3}\chi$. 
]{
\includegraphics[scale=0.4]{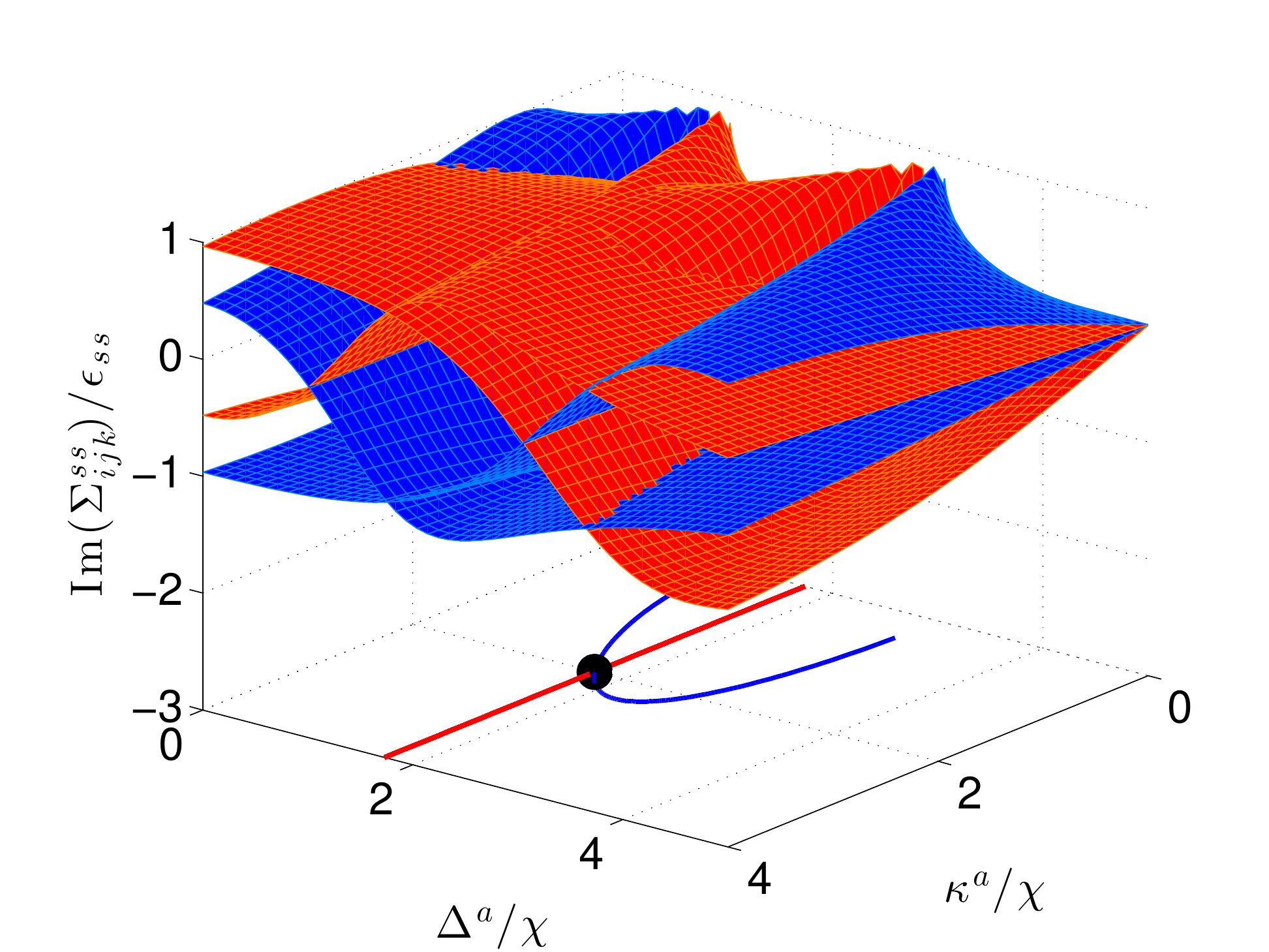}
\label{fig:kernel_a}
}
\subfigure[ The pointer states of the detected field $\Sigma_{ijk}(t)$ for the eigenstates $\ket{000}$ (solid red), $\ket{011}$ (dashed red), $\ket{111}$ (solid blue) and $\ket{001}$ (dashed blue) of the three qubit system. The steady state of the system allows for a perfect parity measurement, while the different transient trajectories result in an undesirable distinguishability within each subspace. The parameters are $\epsilon_0 = \sqrt{\chi}$, $\kappa^a = \kappa^b = 2\chi$, $\Delta^a = -\Delta^b  = \sqrt{3}\chi$ and $\sigma = 10\chi$.  ]{
\includegraphics[scale=0.4]{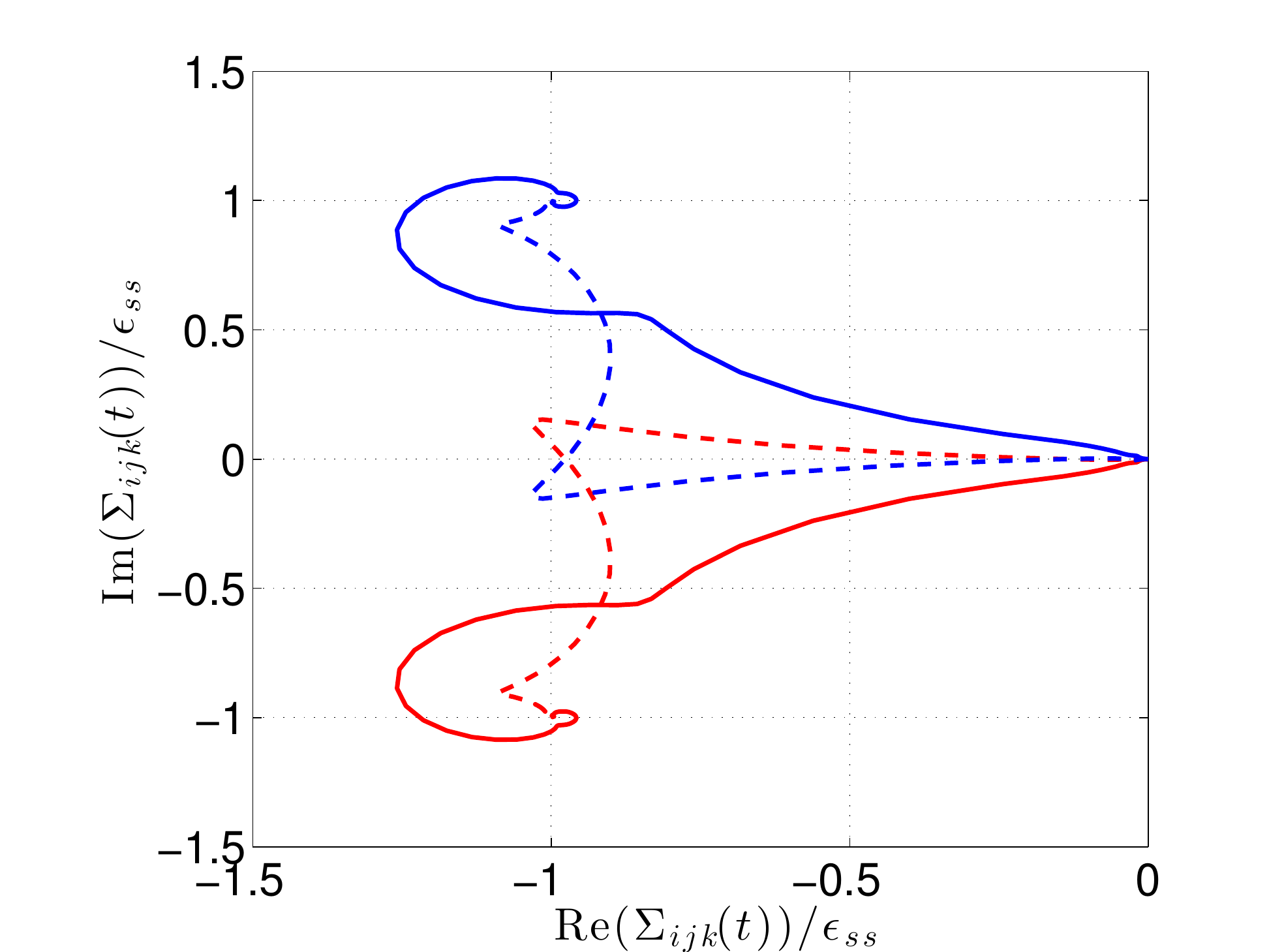}
\label{fig:kernel_b}
}
\caption[Optional caption for list of figures]{The pointer states of measurement.  a) shows the steady state solution $\Sigma_{ijk}^{ss}$ of the pointer states for the different parity subspaces and b) shows the corresponding transient time evolution. }
\label{fig:kernel}
\end{figure}

The goal of an experimental setup as in Fig. \ref{fig:system} is to realize a parity measurement of the joint state of the three qubits, that is, we would like the measurement to distinguish between states belonging to the mutually orthogonal sub-spaces 
\bea
\mathcal{H_+} &=& \mathrm{span}\left(\ket{000}, \ket{011}, \ket{101}, \ket{110}\right), \nonumber \\
\mathcal{H_-} &=& \mathrm{span}\left(\ket{001}, \ket{010}, \ket{100}, \ket{111}\right), 
\eea
without distinguishing between different states within $\mathcal{H_+}$ and $\mathcal{H_-}$. In addition to this, the measurement should not cause any back-action on the measured state apart from the necessary state collapse associated with the gain of information. To realize these properties, the dynamics of the pointer states, together with the chosen measurement basis, must reflect these constraints. In this section we therefore study the evolution given in \Eq{eq:cavity_amplitudes} to obtain a choice of system parameter values such that these conditions are fulfilled. 
We start by analyzing the steady state solution to \Eq{eq:cavity_amplitudes}, given by 
\be\label{eq:sigma_ss}
\Sigma_{ijk}^{ss} = -2 \epsilon_{ss} \frac{\Delta^a\kappa^b + \Delta^b\kappa^a + (\kappa^a + \kappa^b) \chi_{ijk}}{\Delta^b(\kappa^a + \kappa^b) + \Delta^a(\kappa^b + 2i(\Delta^b + \chi_{ijk})) + \chi_{ijk}(\kappa^a + \kappa^b + 2i\chi_{ijk})} \equiv C_{ijk} \epsilon_{ss},
\ee
where $\epsilon_{ss}$ is the steady-state amplitude of the drive and $C_{ijk}$ is a constant which only depends on circuit parameters. Here, the subscripts $ijk$  again refer to the qubit eigenstate $\ket{ijk}$ and we will from now on assume that $\chi_{ijk} = \chi$ is the same for all $i,j,k$, which can be achieved by proper choices of the coupling energies $g_i^j$. In the following we assume the LO phase to be $\phi = \pi/2$ corresponding to a measurement of the imaginary part of the output field $\im(\Sigma_{ijk})$. In order to reduce the complexity of the analysis we limit the number of free parameters by assuming that $\Delta^a = -\Delta^b$ and $\kappa^a = \kappa^b$. In \Fig{fig:kernel_a} we plot $\im(\Sigma)/\epsilon$ as a function of the remaining free parameters $\kappa^a$ and $\Delta^a$. Each surface corresponds to one of the four distinct values of $\chi_{ijk} = \{-3 , -1, 1, 3\}\chi$ which is set by the three-qubit basis states. 
The blue (red) surfaces show the negative (positive) parity solution corresponding to $\chi_{ijk} = \{-1, 3\}(\{1, -3\})\chi$. The intersection between the planes, shown by the blue and red lines on the base of the figure, gives the set of parameter values for which $\im(\Sigma_{000}) = \im(\Sigma_{011})=\im(\Sigma_{101}) = \im(\Sigma_{110}) = \Sigma_+$ and $\im(\Sigma_{111}) = \im(\Sigma_{001})=\im(\Sigma_{100}) = \im(\Sigma_{010}) = \Sigma_-$ such that the states within each subspace cannot be distinguished. As a crucial property, the indicated set of solutions has a symmetry relating the positive and negative parity subspaces.  At the same time, $\Sigma_+ \neq \Sigma_-$, which allows the measurement to distinguish the two subspaces. 

In addition to the conditions imposed by the measurement which are satisfied by the solutions in the blue branch, the solutions in the red branch satisfy the condition that the real parts of the fields also are the same. As discussed in \Sec{sec:eta}, any difference between the pointer states not recorded by the measurement will result in measurement-induced dephasing, so that it is crucial to equate these real parts if we want no additional back-action generated by the measurement. We therefore expect, and numerically find, that deviating from the red, dashed branch decreases the fidelity of the measurement. The indicated point in \Fig{fig:kernel_a} shows the specific choice of parameters used in all numerical simulations which we return to in \Sec{sec:results}. We have numerically verified that changing the values along the red line has only a negligible effect on the fidelity.  \\

The above study has allowed us to extract the right parameter values to achieve the desired properties of the pointer states in the steady state. Equipped with this knowledge we now return to the full solution to \Eq{eq:cavity_amplitudes}. In \Fig{fig:kernel_b}, we plot this as trajectories in the IQ-plane with a pulse shape given by 
\be\label{eq:atanpulse}
\epsilon(t) = \frac{\epsilon_{ss}}{\pi} \left( \arctan\left( \sigma(t - t_{on})  \right) + \frac{\pi}{2} \right).
\ee
%
As expected, the steady state of the solution satisfies the condition that $\im(\Sigma_{ijk})$ is the same for states of the same parity whereas $\Sigma_+ \neq \Sigma_-$. Unfortunately this condition is only valid for the steady state while the transient path taken from the vacuum to the steady state is such that $\Sigma_{000}$
and the fields in the set $\{ \Sigma_{011}, \Sigma_{101}, \Sigma_{110}\}$ can be distinguished by the measurement (as with the pointer states corresponding to states in $\mathcal{H_-}$). This will cause a partial measurement within $\mathcal{H}_\pm$ during this period and therefore decrease the fidelity of the post-measurement state. This is the main source of infidelity of the proposed measurement scheme and the remaining part of the paper is devoted to the study of how to best mitigate this. 
\subsection{Measurement rates}
As discussed above, whenever the measurement is able to distinguish different pointer states from each other, it will give rise to back-action on the qubits. We can quantify the strength of this back-action by considering all the measurement rates and how they correspond to the magnitude of the difference between pointer states. For the specific choice of LO phase made above ($\phi=\pi/2$), the measurement superoperator in \Eq{eq:SMEeff} can be separated into four parts  
\bea\label{eq:explicit_measoperator}
\mathcal{M}[-i\Pi_\Sigma] \rho 
&=& 
\frac{\im(\xi(t))}{2} \mathcal{M}[\Pi_+ - \Pi_-] \rho + 
\frac{\im(\delta(t))}{2} \mathcal{M}[\Pi_{000} - \Pi_{011} - \Pi_{101} - \Pi_{110}] \rho \nonumber \\
&-& 
\frac{\im(\delta(t))}{2} \mathcal{M}[\Pi_{111} - \Pi_{001} - \Pi_{010} - \Pi_{100}] \rho \nonumber \\
&-& 
i\frac{\re(\delta(t))}{2} \left[ \Pi_{000} + \Pi_{111} - \Pi_{001} - \Pi_{010} - \Pi_{011} - \Pi_{100} - \Pi_{101} - \Pi_{110} , \rho \right],
\eea
where we have defined the sum and difference fields $\xi(t) \equiv \Sigma_{000}(t) + \Sigma_{011}(t)$ and $\delta(t) \equiv \Sigma_{000}(t) - \Sigma_{011}(t)$ and used the specific symmetries of the pointer states e.g. $\im(\Sigma_{000}) = -\im(\Sigma_{111})$. The operators $\Pi_\pm = \sum_{ijk \in \mathcal{H}_\pm} \ket{ijk}\bra{ijk}$ are projection operators on $\mathcal{H}_\pm$. The first term in \Eq{eq:explicit_measoperator} represents the gain of information about the parity of the state as expected from the measurement. Hence, we define a parity measurement rate 
\be
\Gamma_m^\mathrm{P}(t) = \eta \im(\xi(t))^2.
\ee
The next two terms arise from the fact that the pointer states within each parity subspace are not identical, resulting in an undesired measurement within each subspace.  This intra-parity subspace measurement rate is given by 
\be
\Gamma_m^\mathrm{IP}(t) = \eta \im(\delta(t))^2.\label{IP}
\ee
The last term gives a stochastic rotation of the phase in the coherences between the subspaces $\mathrm{Span}(\left(\ket{000}, \ket{111}\right)$ and $\mathrm{Span}(\left(\ket{001}, \ket{010}, \ket{100}, \ket{011}, \ket{101}, \ket{110}\right)$, an effect that does not affect the purity of the post-measurement state for a single measurement. It will however affect the purity of the average state. This effect could be cancelled by the use of feedback and poses no fundamental limitation on the measurement fidelity. This cannot, however, be said for the unwanted, intra-parity measurements. Since the two effects are both $\propto \delta(t)^2$, it is clear that we need to make $\delta(t)$ as small as possible to get a measurement with high fidelity. We return to this issue in \Sec{sec:transients}.
\subsection{Measurement fidelity -- two-outcome vs. three-outcome measurement}\label{sec:measfidelity}
In this section we introduce two measures used to assess the fidelity of the measurement; the signal to noise ratio, SNR, and the overlap fidelity of the post-measurement state relative to the pre-measurement state. The SNR quantifies the distinguishability between signals conditioned on states with different parity, while the overlap fidelity measures the undesired back-action the measurement has by comparing the real post-measurement state to the ideal one. \\\\
To convert the time-dependent current into a single measurement outcome we use the integrated current
\be
s_{j(t)}(\tau) = \int_0^\tau j(t) dt
\ee 
as our single (real-valued) measurement result.  Here the measurement time is given by $\tau$. 

We will consider two possible approaches to further interpreting this real-valued outcome as a discrete-valued measurement.  Ideally, the integrated measurement outcome has an unambiguous sign; for some of the measurement parameters considered below, this is in fact the case.  Under these circumstances, it is satisfactory to infer a parity directly from the measurement: $s>0$ meaning even parity, and $s<0$ meaning odd parity.  However, we find that to improve the intra-sector overlap fidelity, it is important to consider measurement parameters that result in a significant number of outcomes with $s\approx 0$.  In this case, it is natural to introduce a finite ``conclusiveness threshold'' $s_{th}$.  
That is, in addition to assigning outcome ``even'' if $s>s_{th}$ and ``odd'' if $s<-s_{th}$, we call the measurement ``inconclusive'' if $|s|<s_{th}$.   A high value of $s_{th}$ allows the observer to discard measurement results that would otherwise lead to a corrupted post-measurement state due to mixing of states with different parity. We will see that allowing a moderate percentage of ``inconclusive'' assignments permits the even/odd overlap fidelity to be dramatically improved in the successful cases. Depending on the objective of the measurement different choices of this threshold will be appropriate, as we discuss further in the Conclusions.  \\\\
For each state $\ket{ijk}$ the current in \Eq{eq:currentq} is given by 
\be\label{eq:current_by_state}
j_{ijk}(t)dt = 2\sqrt{\eta}\im(\Sigma_{ijk}(t))dt + dW(t).
\ee
which, by the linearity of quantum mechanics, gives the current from a general state in $\mathcal{H_\pm}$:  $\ket{\psi_\pm} = \sum_{ijk \in \mathcal{H}_\pm }\gamma_{ijk}\ket{ijk}$
\bea\label{eq:current_pm}
j_\pm(t)dt &\approx& 2 \sqrt{\eta} \sum_{ijk\in\mathcal{H}_\pm} |\gamma_{ijk}|^2 \im(\Sigma_{ijk}(t))dt + dW(t),
\eea
where we have assumed that the coefficients $\gamma_{ijk}$ are unaltered during the measurement, that is, we assume that the transients have negligible effect on the post-measurement state. This assumption can be justified if we consider most of the signal to be generated in the steady state. In the weak-measurement limit discussed in \Sec{sec:transients}, this is a fair assumption. \\

Given a current, $j_\pm(t)$, conditioned on a state in $\mathcal{H}_\pm,$ we define the SNR to be the ratio between the mean and standard deviation of the difference $s_{j_+}(\tau) - s_{j_-}(\tau)$
\be
\mathrm{SNR}(\tau) = 
\frac{\mathrm{E}[s_{j_+}(\tau)] - \mathrm{E}[s_{j_-}(\tau)] }{\sqrt{\mathrm{Var}[s_{j_+}(\tau)] + \mathrm{Var}[s_{j_-}(\tau)]}} 
= \frac{\mathrm{E}[s_{j_+}(\tau)] - \mathrm{E}[s_{j_-}(\tau)] }{\sqrt{2\tau}},
\ee
where we have dropped the time argument in $j_\pm(t)$ for notational transparency and used the statistical properties of the Wiener process in the second equality. 
%
%
Within the assumptions made above, the SNR is given by 
\be
\mathrm{SNR}(\tau) = \sqrt{\frac{2\eta}{\tau}} \int_0^\tau\left( \sum_{ijk\in\mathcal{H}_+} |\gamma_{ijk}|^2 \im(\Sigma_{ijk}(t))  - 
\sum_{ijk\in\mathcal{H}_-} |\gamma_{ijk}|^2 \im(\Sigma_{ijk}(t)) \right)dt
\ee
which can be further approximated if we assume that the fraction of the measurement time spent in the transient region is negligible $\tau \gg 1/\kappa$, that is we make the replacement $\Sigma_{ijk} \to C_{ijk}\epsilon_{ss}$
\be\label{eq:SNR_ideal}
\mathrm{SNR} \approx 2 \sqrt{2} \sqrt{\eta} \im(C_{111}) \epsilon_{ss} \sqrt{\tau}, 
\ee 
where we recall the definition of $C_{ijk}$ below \Eq{eq:sigma_ss}. As expected, the fact that $\mathrm{SNR} \propto \epsilon_{ss} \sqrt{\tau}$ shows that $\mathrm{SNR} \gg 1$ can be achieved for arbitrarily low value of measurement strength $\epsilon_{ss}$. We will further explore this limit in \Sec{sec:transients} when studying the effect of field transients.  \\\\
If one is interested in only measuring the parity of the state it is enough to have $\mathrm{SNR} \gg 1$ for the measurement to be considered high fidelity.  A good example of such a standard, high-fidelity quantum measurement is photon detection using high-fidelity avalanche diodes. Here however the photon is completely destroyed in the process. In a quantum informational setting, the post-measurement state is often to be further processed in some algorithm or error correction scheme. In this case it is crucial that the post-measurement state conditioned on the outcome of the detection is the expected one. As a measure of this we consider the overlap fidelity
\be\label{eq:fidelity_def}
F_\pm =  \sqrt{\bra{\psi_\pm} \mathrm{E}_\pm[\rho]\ket{\psi_\pm} },
\ee  
where $\ket{\psi_\pm}$ is the expected post-measurement state and $ \mathrm{E}_\pm[\rho ]$ is the ensemble averaged, post-measurement state where the mean is taken over states assigned to either $\mathcal{H}_\pm$ by the measurement.

%
\subsection{Measurement efficiency}\label{sec:eta}
\begin{figure}[ht]
\centering
\subfigure[ Measurement results with $\eta = 1$. ]{
\includegraphics[scale=0.4]{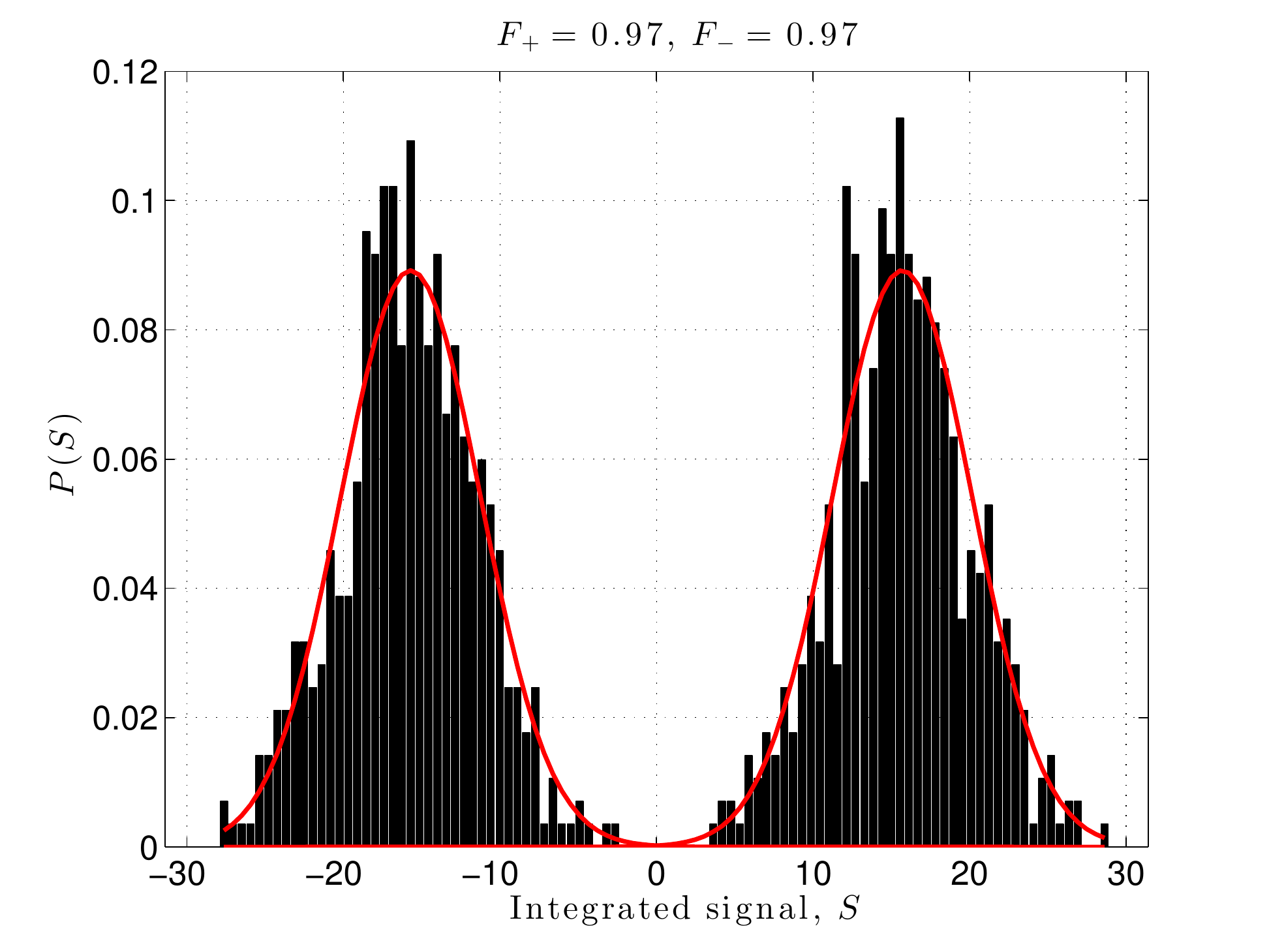}
\label{fig:histograms_eta_a}
}
\subfigure[ Measurement results with $\eta = 0.5$. ]{
\includegraphics[scale=0.4]{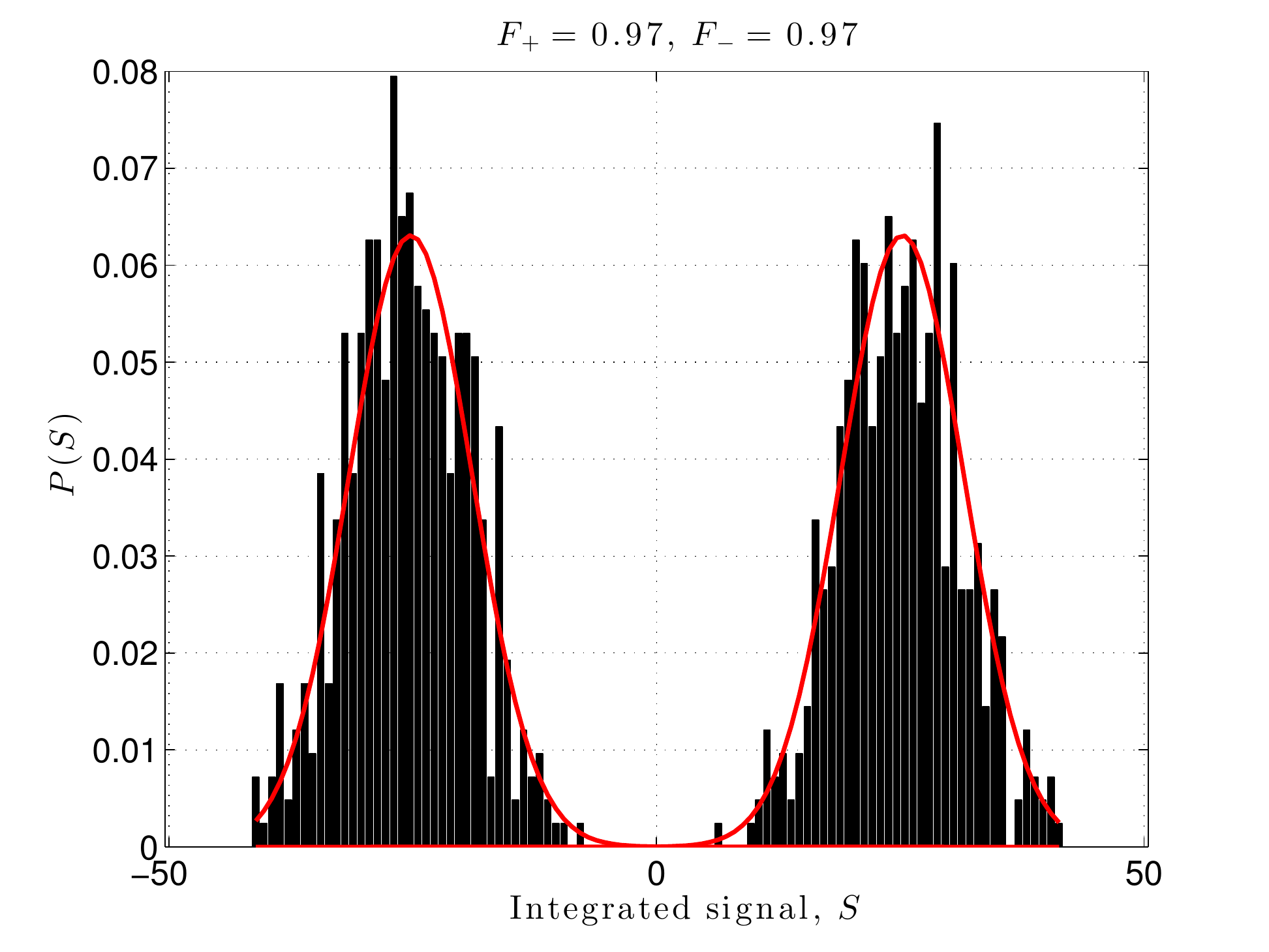}
\label{fig:histograms_eta_b}
}
\caption[Optional caption for list of figures]{The effect of measurement efficiency. Histograms of 1000 measurement results corresponding to the pre-measurement state in \Eq{eq:psi_pre}. The parameters are as in \Fig{fig:kernel} with $\gamma_{1j} = \gamma_{\varphi j} = \gamma_p = 0$.  The measurement time and drive strength are chosen to be  $\tau = 20/(\eta\chi)$, $\epsilon_{ss} = 2\sqrt{\chi/20}$ corresponding to $\mathrm{SNR} = 4\sqrt{2}$ in both cases. The value of $\eta$ is specified in each subfigure. The red curves are normal distributions with mean and variance defined in \Eq{eq:gaussian}. The values of $F_\pm$ are given above each figure.}
\label{fig:histograms_eta}
\end{figure}
The quantum efficiency, $\eta$, quantifies how much of the information, which is flowing out of the system, is actually measured. Given a pure initial state and a quantum limited measurement, that is no additional back action apart from the necessary state collapse, with $\eta = 1$ the projection postulate ensures that the post-measurement state is pure. For $\eta < 1$ this is in general no longer true since the observer must average over the non-observed measurement results to obtain the post-measurement state. This procedure is the origin of measurement induced dephasing and lowers the fidelity of the measurement in the sense of \Eq{eq:fidelity_def}. In \Fig{fig:histograms_eta}, we plot the histograms corresponding to 1000 measurement results for $\eta = 0.5$ and $1$.  In both cases the initial state is given by 
\be\label{eq:psi_pre}
\ket{\psi}_\mathrm{pre} = \frac{1}{\sqrt{8}}\sum_{ijk}\ket{ijk},
\ee 
which for a perfect parity measurement would be projected on to the post-measurement states
\bea\label{eq:psi_post}
\ket{\psi}_+ = \frac{1}{\sqrt{4}}\left( \ket{000} + \ket{011} + \ket{101} + \ket{110} \right), \nonumber \\
\ket{\psi}_- = \frac{1}{\sqrt{4}}\left( \ket{001} + \ket{010} + \ket{100} + \ket{111} \right),
\eea
with equal probability. While confining our attention to this initial state does not explore all aspects of the measurement superoperator, it is optimally sensitive to any loss of intra-sector coherence during the measurement, and it is a state with a structure, with its equal superposition of qubit basis states, resembling that of the important stabiliser states of quantum error correction\cite{NC}. 

The measurement drive strength $\epsilon_{ss}$ is kept the same for the two cases in \Fig{fig:histograms_eta}. The SNR is also held at a constant value by increasing the measurement time to compensate for the lower value of $\eta$. Since we are interested in the effect of lowering $\eta$, we ignore the effect of decoherence, that is, $\gamma_{1j} = \gamma_{\varphi j} = \gamma_p = 0$ where 
\begin{equation}
\gamma_ p = (g/\Delta)^2(\kappa^a + \kappa^b) \label{exPurcell}
\end{equation}
is the Purcell decay rate. 
The red curves in \Fig{fig:histogram} are normal distributions with mean and variance
\be\label{eq:gaussian}
\mathrm{E}[s_\pm(\tau)] = \mp 2\sqrt{\eta}\int_0^\tau \frac{1}{4}\im\left( \Sigma_{000}(t) + 3\Sigma_{011}(t) \right) dt, \qquad
\mathrm{Var}[s_\pm(\tau)] = \sqrt{\tau}.
\ee
\\
From the overlap fidelity, it is clear that the purity of the state is not affected by the decrease in $\eta$. This robustness comes from the fact that the pointer states corresponding to states within $\mathcal{H}_\pm$ are perfectly indistinguishable in the steady state. Hence there are no unrecorded measurement results to average over and the state remains pure. The fact that $F_\pm < 1$ is an effect of the transient evolution of the pointer states which is not affected by the measurement efficiency. Note however that, in the presence of decoherence, the measurement efficiency will have an indirect effect on $F_\pm$ through the longer measurement times needed to keep SNR high.
\section{Effect of field transients}\label{sec:transients}
\begin{figure}[ht]
\centering
\subfigure[ Measurement results for short measurement time, $\tau = 10/\chi$ and $\epsilon_{ss} = 2/\sqrt{\tau}$.]{
\includegraphics[scale=0.4]{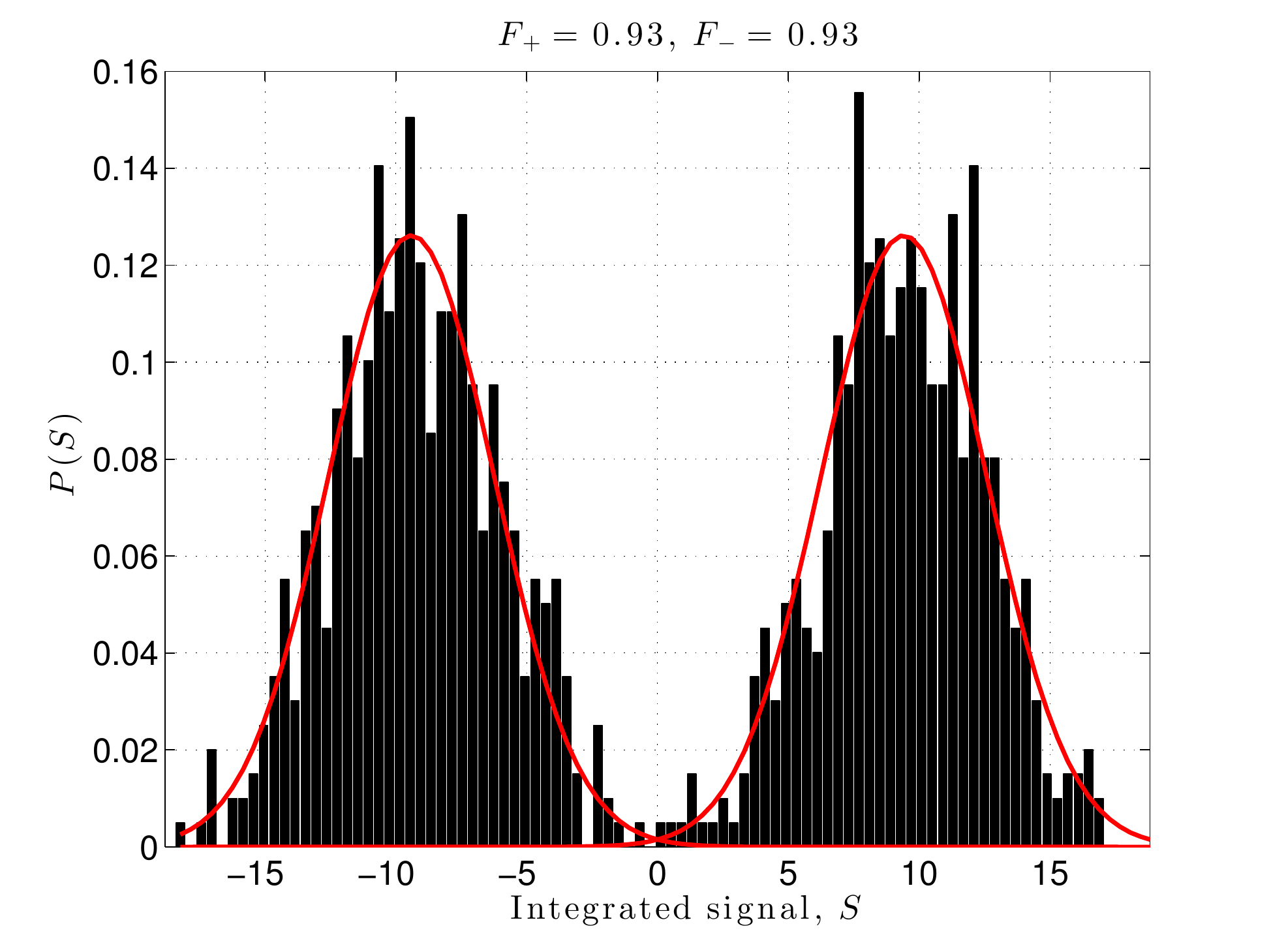}
\label{fig:histograms_a}
}
\subfigure[ Measurement results for long measurement time, $\tau = 100/\chi$ and $\epsilon_{ss} = 2/\sqrt{\tau}$.]{
\includegraphics[scale=0.4]{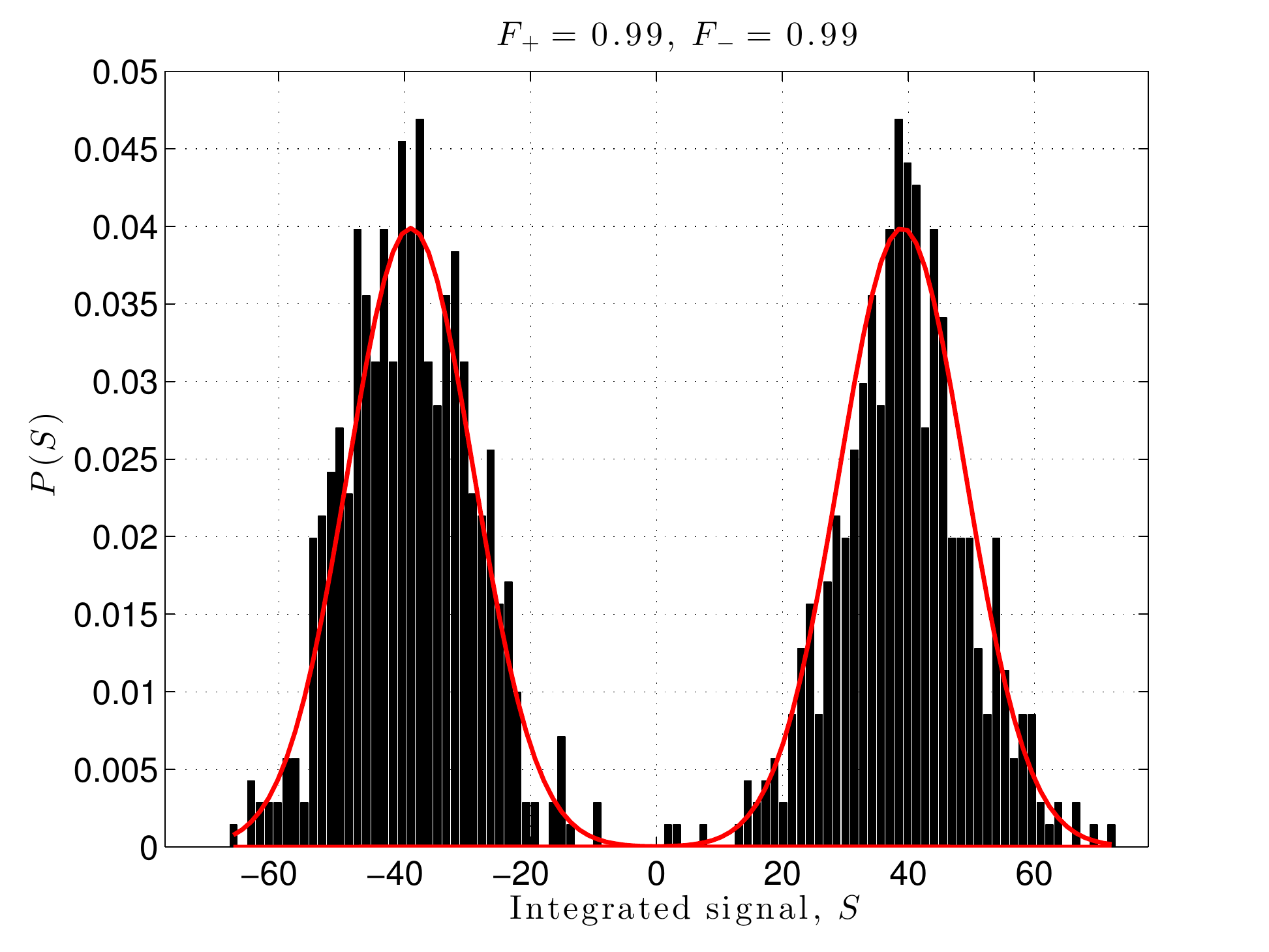}
\label{fig:histograms_b}
}
\caption[Optional caption for list of figures]{The effect of transients. Histograms of 1000 measurement results corresponding to the pre-measurement state in \Eq{eq:psi_pre} and $\mathrm{SNR} = 4\sqrt{2}$. The parameters are as in \Fig{fig:kernel} with $\gamma_{1j} = \gamma_{\varphi j} = \gamma_p = 0$. The values if $\epsilon_{ss}$ and $\tau$ are specified in each subfigure. The red curves are normal distributions with mean and variance defined in \Eq{eq:gaussian}. The values of $F_\pm$ are given above each figure.}
\label{fig:histogram}
\end{figure}

It is clear that the integrated rate $\Gamma_m^\mathrm{IP}(t)$ of Eq. (\ref{IP}) (total effect on the state) must be minimized to limit the unwanted effect due to the transient behavior of the pointer states. 
%
%
To realize this we make the observation that $\delta(t)\to 0$ when $\epsilon_{ss} \to 0$. This is also true for $\xi(t)$ and the measurement will therefore be weak in this sense. The measurement can however still be strong in the sense that the SNR defining product $\epsilon_{ss}\sqrt{\tau}$ can be kept constant by increasing the measurement time such that the value of this product is kept constant as $\epsilon_{ss} \to 0$. In the absence of decoherence mechanisms, we can keep the measurement on for as long as we want, and in this way realize a strong measurement while mitigating the effect of the unwanted back action. In \Fig{fig:histogram}, we plot the histograms of 1000 measurement results for two different values of $\epsilon_{ss}$ and $\sqrt{\tau}$ such that $\epsilon_{ss} \sqrt{\tau} =  2$ but varying $\tau$. 

The initial state is given in \Eq{eq:psi_pre} and, since we want to single out the detrimental effect of the field transients, we ignore the effect of decoherence, that is, $\gamma_{1j} = \gamma_{\varphi j} = \gamma_p = 0$. The histograms in \Fig{fig:histogram} show comparable overlap for the different measurement strengths, but the overlap fidelity is near unity for the long measurement time $\tau = 100/\chi$ while being significantly lower for the shorter measurement time $\tau = 10/\chi$. This clearly shows that, in the limit of long measurement time, that is, weak measurement pulse, the setup considered here makes for a perfect parity measurement. Although the SNR defining product $\epsilon_{ss}\sqrt{\tau}$ is chosen to be identical for both measurement times, we note that this only estimates the SNR accurately if all of the measurement takes place in the steady state. For $\tau = 10/\chi$, we are approaching the limit where the transient behavior makes up a non-negligible part of the measurement duration, leading to lower SNR. \\%
%

\begin{figure}[ht]
\centering
\subfigure[ The drive pulse $\epsilon(t)$ for different values of rise time $\sigma$ and the corresponding difference $\im(\delta(t))$. ]{
\includegraphics[scale=0.4]{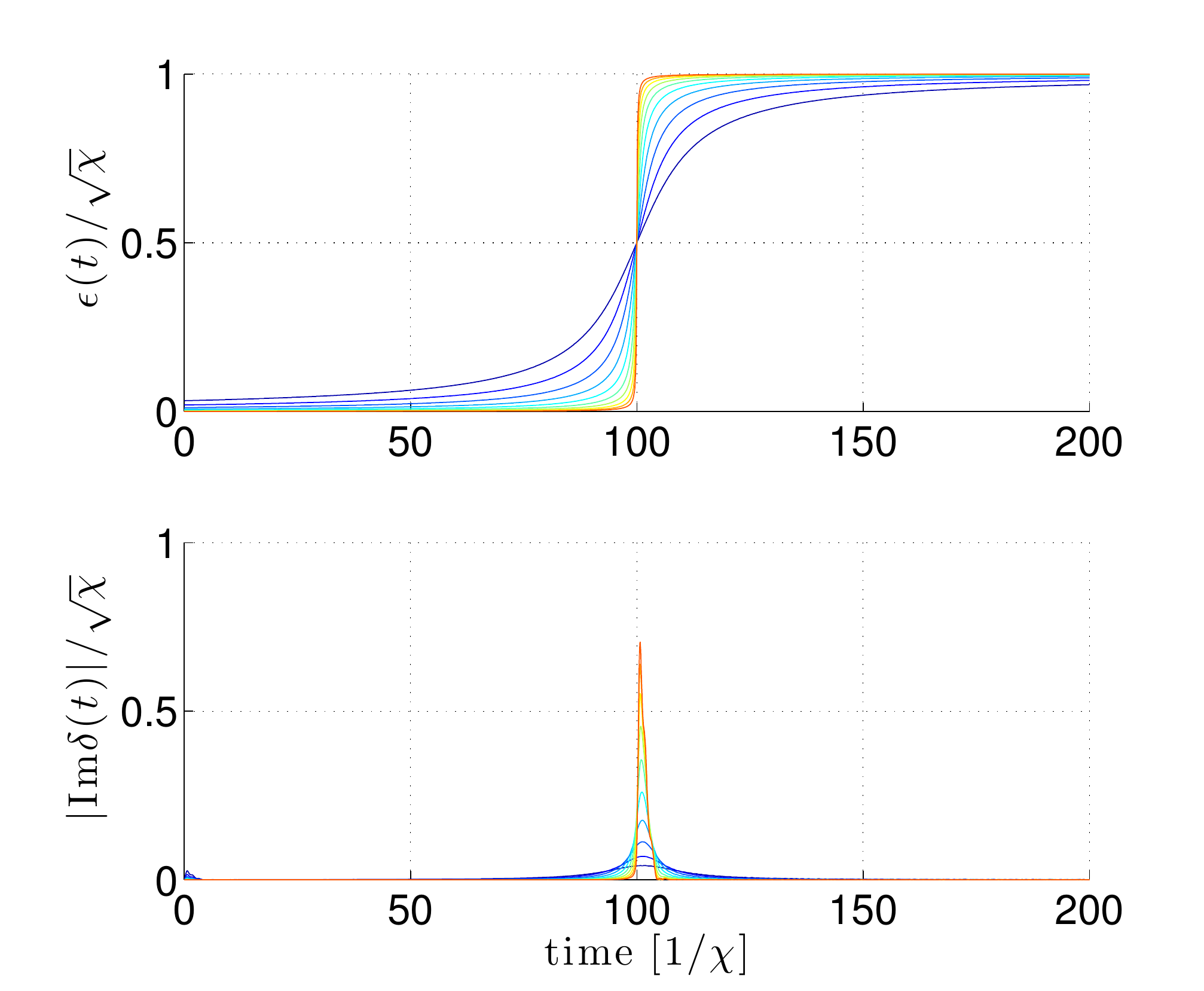}
\label{fig:field_differences_a}
}
\subfigure[ The integrated difference for a large range of rise times $\sigma$. The effect is on the order of a few percent. ]{
\includegraphics[scale=0.4]{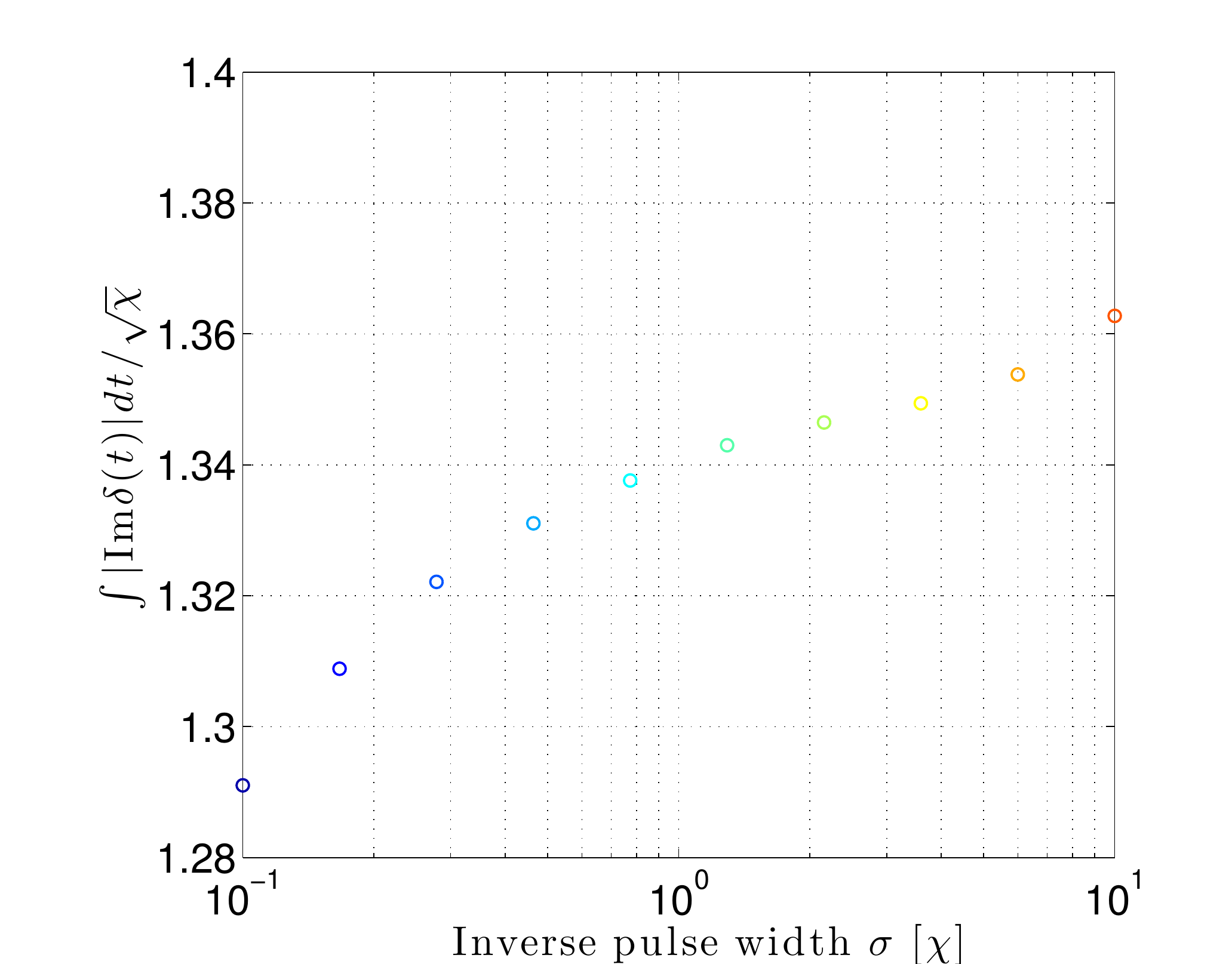}
\label{fig:field_differences_b}
}
\label{fig:field_differences}
\caption[Optional caption for list of figures]{ The difference field $\delta(t)$ and $\int \delta(t) dt$ for a pulse given in \Eq{eq:atanpulse}. }
\end{figure}

Another way to possibly enhance the measurement fidelity is to have the measurement pulse $\epsilon(t)$ turned on slowly compared to all the other time scales in the system. This allows the pointer states to approximately follow the paths given by their instantaneous steady state value. We consider the pulse shape given \Eq{eq:atanpulse} and plot $|\im(\delta(t))|$ in \Fig{fig:field_differences_a} for a large range of rise times $1/\sigma$. 
As expected, the sharper the onset of $\epsilon(t)$ is, the bigger the difference $\delta(t)$ becomes between the pointer states in the same parity subspace. Making the pulse smoother in time decreases this difference but simultaneously spreads it over a larger time. To quantify the effect of this trade-off, we plot the integrated value of $\im(\delta(t))$ in \Fig{fig:field_differences_b}. We see that the difference is negligible over a large range of $\sigma$. We can therefore safely say that the measurement time will not be limited by any adiabaticity constraints and, in the remainder of this work, we focus on the effect of long measurement time as described in the beginning of this section.

\section{Results for optimal measurement}\label{sec:results}
So far all the results have been derived without considering the detrimental effect of decoherence, which is inevitablly present due to coupling of the cavity modes to the continuum.   From the analysis so far it is however evident that, for the measurement to be high fidelity, we need a long measurement time such that $\epsilon_{ss} \sqrt{\tau} \gg 1$ and $\epsilon_{ss} \to 0$. But increasing $\tau$ indefinitely is not possible in the presence of qubit decay mechanisms. Thus, there exists an optimal measurement time $\tau_\mathrm{opt}$, which we will identify below, for which the two competing effects of transients and qubit relaxation balance one another.   In this optimisation for $\tau_\mathrm{opt}$ we will fix the SNR given in \Eq{eq:SNR_ideal} at a desired value and calculate the measurement record and post-measurement state for different measurement times. Note that the actual calculated SNR will differ from the estimate that we used to fix the relationship between time and drive strength. This is due to the fact that the expression in \Eq{eq:SNR_ideal} is an idealization -- decoherence will cause additional dynamics not accounted for in that analysis. \\

\begin{figure}[htp] 
\centering
\includegraphics[scale=0.5]{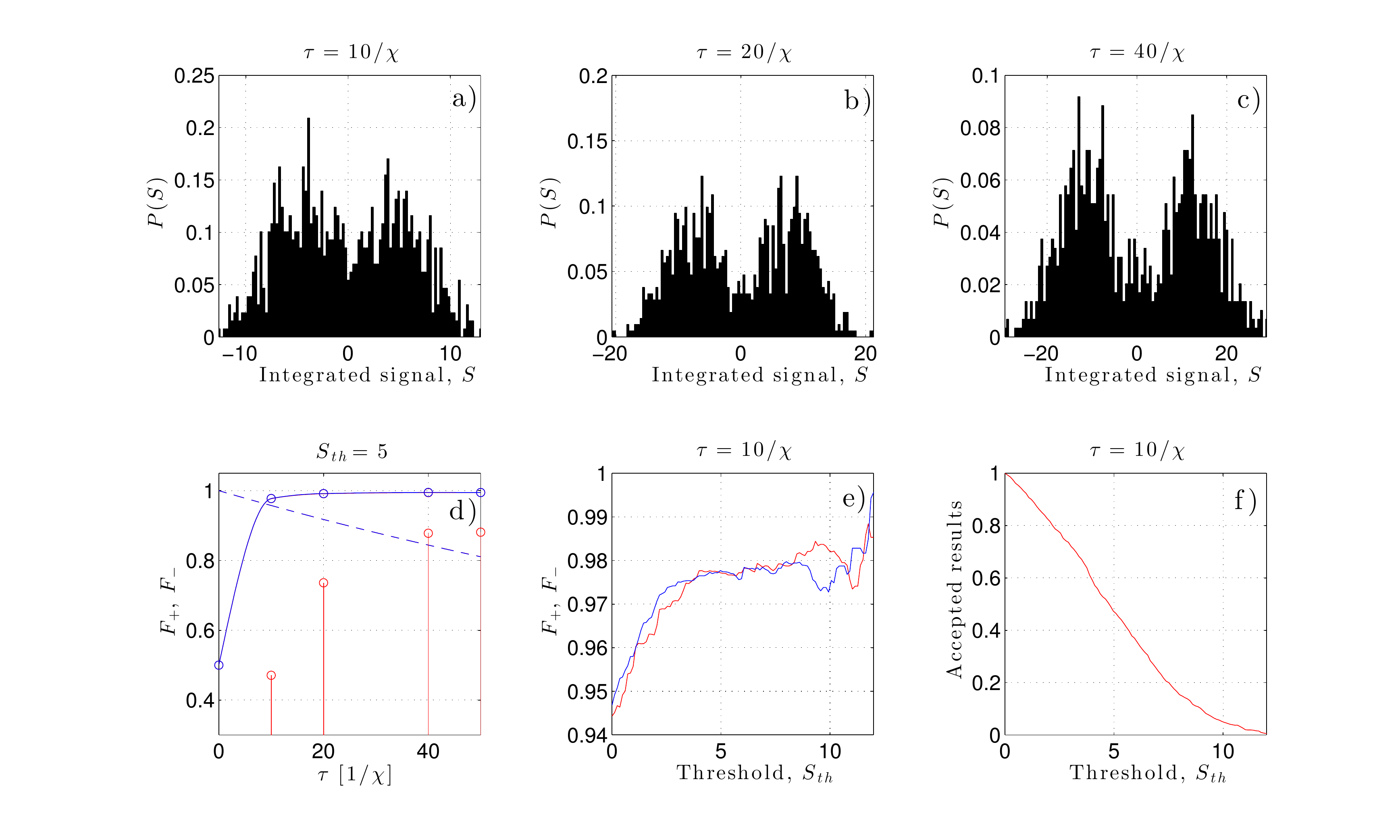}
\caption{Measurement results for $\eta=1$ without decoherence, compared with simple Purcell model.  Histograms a), b), and c) show the measurement results $s(\tau)$ with a fixed SNR = $2\sqrt{2}$. In d) we plot the overlap fidelity as function of measurement time (overlapping solid blue and red) along with the fraction of accepted results (red stem). The overlapping blue and red dashed lines show $F_\pm$ in the absence of measurement, including the effect of decoherence, with $\gamma_p = \chi/400$ and $\gamma_\varphi=\chi/300$.  In e) we plot $F_\pm$ as a function of $s_{th}$ for $\tau = 10/\chi$ and f) shows the corresponding fraction of accepted measurement results.}
\label{fig:no_purcell}
\end{figure}

In \Fig{fig:no_purcell} we plot the integrated measurement record and the overlap fidelity of the corresponding post-measurement state for a relatively small SNR = $2\sqrt{2}$. The initial state is given in \Eq{eq:psi_pre}. Here we have not included any decoherence effects but the objective is rather to see how good the measurement can be in the absence of imperfections, just taking into account the effect of field transients. 
Each histogram in Fig. \ref{fig:no_purcell}(a), (b), and (c) shows the measurement results for 1000 simulated trajectories. In \Fig{fig:no_purcell}(d), we plot the overlap fidelities $F_\pm$, which are essentially identical for even and odd parities, along with the fraction of conclusive measurement results (even plus odd) for $s_{th} = 5$. As the measurement time is increased, the fidelity approaches unity yielding a perfect parity measurement. The fraction of accepted measurement results also increase with $\tau$ as the SNR approaches that of \Eq{eq:SNR_ideal}.\\
 
The dashed lines in \Fig{fig:no_purcell}(d) (essentially identical for even and odd) are the overlap fidelities without measurement, only including the effect of decoherence. Here, the initial states are given by $\ket{\psi_\pm}$. These lines give a benchmark for how fast we need to perform the measurement in order to not be limited by decoherence. Here, we choose the Purcell rate $\gamma_ p = \chi/400$ and dephasing $\gamma_\varphi = \chi/300$ such that if $\chi = 1$MHz, the relaxation rate and decoherence rate would be $T_1 = 1/\gamma_p = 400 \mu s$ and $T_2^* = 1/(\gamma_p/2 + \gamma_\varphi) = 218 \mu s$ respectively. Such a value for $\gamma_\varphi$ has been obtained in 3D circuit QED architectures \cite{RigettiPRB2013}, while the value of $\gamma_p$ is $\sim 4$ times smaller than current state-of-the-art experimental values (note that we have included all contributions to relaxation into $\gamma_p$ for simplicity). We believe that, with the ongoing experimental progress in improving these numbers \cite{SandbergAPL2013, ChangAPL2013}, numbers like the above should be possible in the near future. It should be noted that any application with the need for multiple qubits would place similar requirements on longer qubit lifetimes. For this choice of parameter values, we see that the measurement time needs to be $\tau \simeq 10/\chi$ since, for larger times, the measurement is limited by decoherence. \\

In \Fig{fig:no_purcell}(e) we plot $F_\pm$ as a function of $s_{th}$ for a measurement time of $\tau = 10/\chi$. By discarding measurement results (i.e., labelling them inconclusive), we can increase the conditional fidelity of the post-measurement state up to $> 98\%$. Note that for $s_{th} > 10$, the number of accepted measurement results are too few to allow good statistics, hence the increase in variance of $F_\pm$. The decreasing fraction of accepted results as function of $s_{th}$ is plotted in \Fig{fig:no_purcell}(c).   Note that, we can get estimates for the physical parameters implied by these parameter settings: ignoring the distinction between different modes and different qubits, we get, using Eq. (\ref{exPurcell}) and the standard dispersive relation $\chi=g^2/\Delta$,
\begin{equation}
\Delta=4\chi^2/\gamma_p,\,\,\,\,g=\sqrt{4 \chi^3/\gamma_p}.
\end{equation}
This gives numerical values $\Delta= 1.6$GHz and $g=40$MHz.  We see that a large value of detuning, combined with a moderate value of the qubit-cavity coupling constant $g$, gives the best measurement.  Note that in order to avoid direct qubit-qubit coupling, the detuning $\Delta$ should be different from one qubit to the other by, say, hundreds of MHz, with the $g$'s correspondingly adjusted so that the $\chi$ parameters are all equal.

\begin{figure}[htp] 
\centering
\includegraphics[scale=0.5]{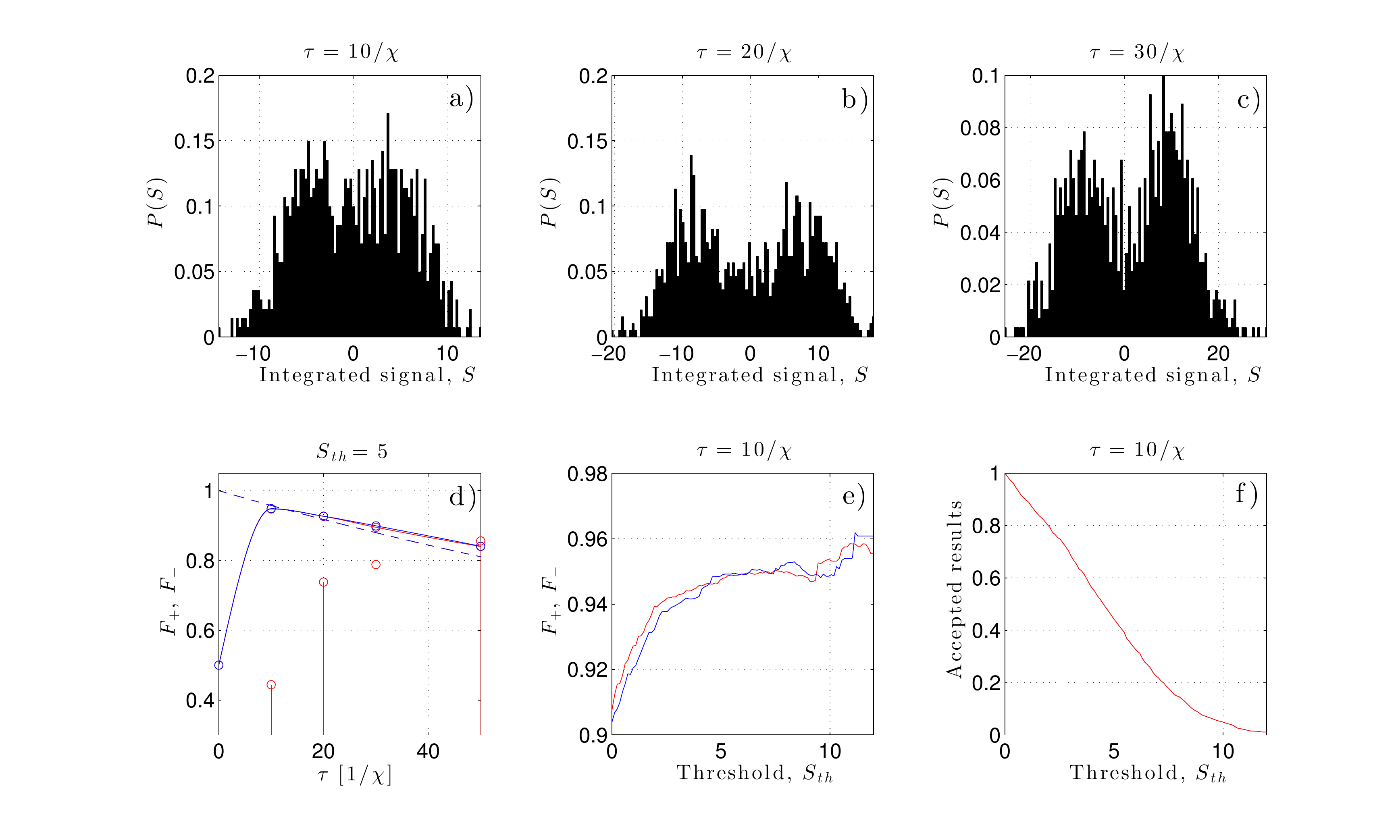}
\caption{Measurement results with decoherence. Histograms a), b), and c) show the measurement results $s(\tau)$ with a fixed SNR = $2\sqrt{2}$. In d) we plot the overlap fidelity as function of measurement time (overlapping solid blue and red) along with the fraction of accepted results (red stem). The overlapping blue and red dashed lines show $F_\pm$ in the absence of measurement, including the effect of decoherence. In e) we plot $F_\pm$ as a function of $s_{th}$ for $\tau = 10/\chi$ and f) shows the corresponding fraction of accepted measurement results. The parameters are as in \Fig{fig:kernel} with $\gamma_p = \chi/400$, $\gamma_\varphi = \chi/300$ and $\eta =1$.}
\label{fig:with_Purcell}
\end{figure}

In \Fig{fig:with_Purcell} we plot the same quantities as in \Fig{fig:no_purcell} but with decoherence included. 
The peaks in the histograms are less separated than in \Fig{fig:no_purcell} since the Purcell decay mixes the different parity subspaces. The overlap fidelity in \Fig{fig:with_Purcell}(d) follows the fidelity set by the decoherence in absence of measurement (dashed lines). We observe that the fidelity is actually slightly better with the measurement on, which we can understand as a type of Zeno-effect. Since the Purcell relaxation is dominated by single qubit bit-flip errors, and these also change the parity of the state, the measurement partly protects the state from the dominant decay process with higher fidelity as a consequence. From \Fig{fig:with_Purcell}(d) we see that a post-measurement overlap fidelity of $\sim 90 \%$ is possible which can be increased to $\sim 95\%$ provided that $\sim 60\%$ of the measurement results are labelled as inconclusive. 
%
%
\section{Conclusions}\label{sec:conclusions}
In conclusion, we have performed an analysis of a three qubit parity measurement in a circuit-QED setup where the joint state of two single mode resonances are used as pointer states of the measurement. We find that the measurement fidelity is limited by the transient dynamics of these pointer states and show that this limitation, in the absence of other decoherence mechanisms, can be overcome by the use of a weak probe signal. In this limit we can still obtain a high signal-to-noise ratio due to the fact that the steady state of the pointer states perfectly fulfills the conditions of a parity measurement, and the weak probe can thus be compensated by a longer measurement time. In the presence of additional decay, this strategy breaks down; but we show that, with realistic numbers for the decoherence, we can obtain a state fidelity of $\sim 95\%$ for the post-measurement state provided that we throw away $\sim 60\%$ of the measurement results.  \\

The sort of measurement described here has direct application to the implementation of fault tolerant quantum computation using topological error correction codes employing three-qubit checks in the code of \cite{3par}, or the analogous four-qubit parity checks in the surface code\cite{BT}.When such applications are attempted, it is clear that very different statistical considerations would be brought to bear in the interpretation of the measurement results.  In such error correction, there is an ``error free'' state of fixed parity, and the measurement is expected to give this outcome a large majority of the time. At a minimum, this would, on account of Bayesian reasoning, move the threshold $s_{th}$ away from its symmetric setting.  Furthermore, there would never be any reason to interpret any measurement outcome as 100\% conclusive, since optimal corrective actions will be inferred from a large amount of measurement data of varying degrees of certainty. Finally, the correlation of measurement outcome $s$ with the overlap fidelity $F$ changes the interpretation of subsequent error syndrome measurements, because a departure of $F$ from unity implies a degradation of the multiqubit state which will be expected to show up as an erroneous parity outcome in the conjugate basis, which is needed on overlapping clusters of qubits in the surface code.  More research will be needed to determine what measurement SNRs and fidelities are needed for the topological error correction to be successful. \\

Of course, there are further problems that are untouched by the present analysis; most real qubits have more than two quantum levels, which requires an extension of the present analysis, and brings in the possibility of leakage out of the computational space.  It is only beginning to be understood how to effectively deal with leakage-type errors in surface-code error correction.\cite{leak}  Nevertheless, the fact that there is no fundamental limitation to the fidelity of the proposed measurement scheme, indicates that as qubits with yet longer decoherence times become available, our circuit QED-based measurement schemes will become a prime tool for the preservation and control of complex quantum-computational states.  \\

\section*{Acknowlegements}
We thank Firat Solgun and Ben Criger for useful discussions.  DDV and SB are grateful for support from the Alexander von Humboldt
foundation.  LT acknowledges financial support from the Swedish Research Council, and the EU through the projects SOLID and ScaleQIT.

\end{document}